\documentclass{article}
\usepackage{graphicx}
\usepackage{geometry}
\usepackage{placeins}
\usepackage{xcolor}
\usepackage{hyperref} 
\usepackage{subcaption}
\usepackage{textgreek}
\usepackage[OT1]{fontenc}
\graphicspath{FFig}
\title{Lacking oceanic-driven internal multidecadal climate variability is compensated by forced variability in model simulations}
\author{Rapha\"el H\'ebert \& Thomas Laepple}
\begin{document}
\tableofcontents

\newpage
\maketitle
\section{Main Text}
\subsection{Abstract}
Regional climate change in the 21\textsuperscript{st} century will result from the interplay between human-induced changes and internal climate variability. Competing effects from greenhouse gas warming and aerosol cooling have historically caused multidecadal forced climate variations overlapping with internal variability. Despite extensive historical observations, disentangling the contributions of internal and forced variability remains debated, largely due to the uncertain magnitude of anthropogenic aerosols.

Here, we show that, after removing CO\textsubscript{2}-congruent variability, multidecadal temperature variability in instrumental data is largely attributable to internal processes of oceanic origin. This follows from an emergent relationship, identified in historical climate model simulations, between the driver of variability in oceanic regions and the land-ocean variance ratio in the mid-latitudes. Thus, climate models with higher residual (non-CO\textsubscript{2}) forced variability, largely linked to volcanic and anthropogenic aerosols, exhibit more spatially coherent and amplified temperature patterns over land compared to observations. In contrast, models with higher internal variability agree better with the instrumental data. Our results underscore that internal modes of ocean-driven variability may be too weak in many climate models, and that current projections may be underestimating the range of internal variability in regions with high oceanic influence.

\subsection{Forced and internal slow variability}
Projections of regional climates in the 21\textsuperscript{st} century require climate models to accurately simulate both natural and anthropogenically forced variability. Although uncertainty remains regarding the amplitude and spatial pattern of anthropogenically forced variability\cite{hebert_regional_2018}, this component is better understood and can be projected on multidecadal timescales based on emission pathways\cite{intergovernmental_panel_on_climate_change_near-term_2014}. For regional climates, natural variability plays a larger role in projection uncertainty\cite{deser_uncertainty_2012}, although model results suggest that it could still be relatively small\cite{lehner_partitioning_2020}. It is unclear however whether models realistically simulate natural variability. Paleoclimate evidence suggests that models underestimate the amplitude of supra-decadal variability\cite{laepple_ocean_2014,hebert_millennial-scale_2022,laepple_regional_2023}. This has created a discrepancy: models appear to capture the continuum of variability in global mean temperatures\cite{zhu_climate_2019}, mainly driven by external forcing, but not the regional one\cite{laepple_ocean_2014,hebert_millennial-scale_2022}. To address this issue, it has recently been proposed that the spatial covariance structure in models may be a key factor and that regional modes of variability should be more persistent in space and in time\cite{laepple_regional_2023}. 
However, the mechanisms behind a more persistent spatial structure remain unclear, particularly whether they are primarily linked to internal variability or external forcing.

The relative contributions of forced and internal variability in generating multidecadal temperature variability (MDV) have long been debated. Oceanic internal processes, particularly those related to the Atlantic meriodional overturning circulation (AMOC), have traditionally been viewed as the main drivers of Atlantic MDV\cite{delworth_interdecadal_1993,latif_natural_2022}, and in turn of the MDV of the northern hemisphere\cite{zhang_can_2007}. More recently, however, model experiments have suggested that a combination of anthropogenic and volcanic aerosols play a dominant role in driving Atlantic MDV over the historical period\cite{booth_aerosols_2012,bellomo_historical_2018,mann_absence_2020} and the last millennium\cite{mann_multidecadal_2021}, relegating internal variability to a secondary role. 

This hypothesis remains contentious, as significant discrepancies between models and observations have been noted with respect to ocean heat content, surface heat fluxes and also multidecadal sea-surface temperature patterns\cite{zhang_have_2013,kim_key_2018,li_investigation_2020,latif_natural_2022}. Nevertheless, comprehensive attribution studies have shown that anthropogenic aerosols play a key role in the strengthening of the AMOC found in CMIP6 historical simulations over the second part of the 20\textsuperscript{th} century\cite{robson_role_2022,hassan_anthropogenic_2021}, but also noted that excessive aerosol forcing leads to incompatibilities with observations\cite{robson_role_2022,hassan_anthropogenic_2021}. This echoes previous studies supporting that observational constraints imply less negative aerosol forcing than currently implemented in most models\cite{stevens_rethinking_2015,hebert_observation-based_2021,albright_origins_2021}, and that the CMIP6 models with the weakest aerosol forcing tend to show less mid-century cold bias\cite{flynn_strong_2023}. 

While aerosol forcing has improved the alignment between observed and simulated global mean surface temperature—particularly in explaining the post-war cooling and post-1990 warming\cite{stott_observational_2006,booth_aerosols_2012,haustein_limited_2019,wilcox_influence_2013,hebert_observation-based_2021}—its magnitude may be overestimated\cite{stevens_rethinking_2015}. A higher aerosol forcing could be offset by a higher sensitivity to greenhouse gases to yield a realistic global response, and a (potentially overfitted) Atlantic MDV in phase with observations\cite{stott_observational_2006,chylek_atlantic_2014}. Reducing the still large uncertainties around aerosol forcing is critical\cite{bellouin_bounding_2020}, as the resulting attribution could either imply that the Paris agreement targets are attainable, or very challenging\cite{watson-parris_large_2022}.

The study of global mean temperature alone is insufficient to draw conclusions about the relative magnitude of internal and forced variability. Spatially heterogeneous responses may offset one another, and the inclusion of sub-global information can have a significant impact on attribution studies\cite{stott_observational_2006}. The spatial imprint and modes of internal climate variability on longer than decadal timescales have been investigated in both instrumental and model data\cite{brown_regions_2015,cheung_comparison_2017,parsons_magnitudes_2020,wills_pattern_2020}. Studying internal variability from observational data though requires additional assumptions to separate it from forced variability, and these assumptions can influence the spatial patterns obtained\cite{brown_regions_2015,kravtsov_comment_2017}. 

Despite these challenges, the magnitude of multidecadal internal variability was generally found to be insufficient to match observational estimates for the global mean surface temperature\cite{brown_regions_2015}, as well as sea-surface temperature variability in key regions like the northern Pacific and Atlantic oceans\cite{cheung_comparison_2017}. Parsons et al.\cite{parsons_magnitudes_2020} identified distinct spatial imprint of forced and unforced variability in global mean temperatures, but did not attempt to disentangle the two in the instrumental data. A promising method, low-frequency component analysis, has been used to identify spatially coherent slow modes of variability in observations based on timescale separation\cite{wills_pattern_2020}, showing the existence of multidecadal internal variability independent from global warming in the North Pacific\cite{wills_disentangling_2018} and North Atlantic\cite{wills_oceanatmosphere_2019}.

To further investigate the underlying mechanisms, this study examines the spatial patterns of MDV in surface air temperature over both land and ocean using instrumental data (100-member HadCRUT5 ensemble\cite{morice_updated_2021}) and an extensive multi-model ensemble (hereafter models for simplicity) of CMIP6 climate models (9 models with 30 simulations per model, Supp. Table~\ref{tab:cmip6_models}, Methods). By analyzing the distinct spatial covariance structure and land-ocean contrast of forced and internal variability on different timescales, we better constrain the relative contribution of each component.

\subsection{Global to local spatial patterns of multidecadal variability}
Studying internal multidecadal climate variability (MDV) in the recent period is challenging due to its overlap with anthropogenic forcing, which has become dominant even at regional scales. For climate model simulations, the internal variability $T_{Internal}(\overrightarrow{x},t)$ can be well approximated by subtracting the single-model ensemble mean, given a large enough ensemble of simulations under the same boundary conditions\cite{kravtsov_comment_2017}. However, for instrumental data, disentangling internal variability is less straightforward and requires assumptions. One common approach is to estimate forced variability from a multi-model ensemble and subtract it from observations\cite{frankcombe_separating_2015}, but this method can introduce errors, relative to the true (and unknown) forced variability, and generate spurious internal variability and misleading spatial patterns\cite{brown_regions_2015,kravtsov_comment_2017}. 

\begin{figure}[ht]
\includegraphics[width=0.6\linewidth]{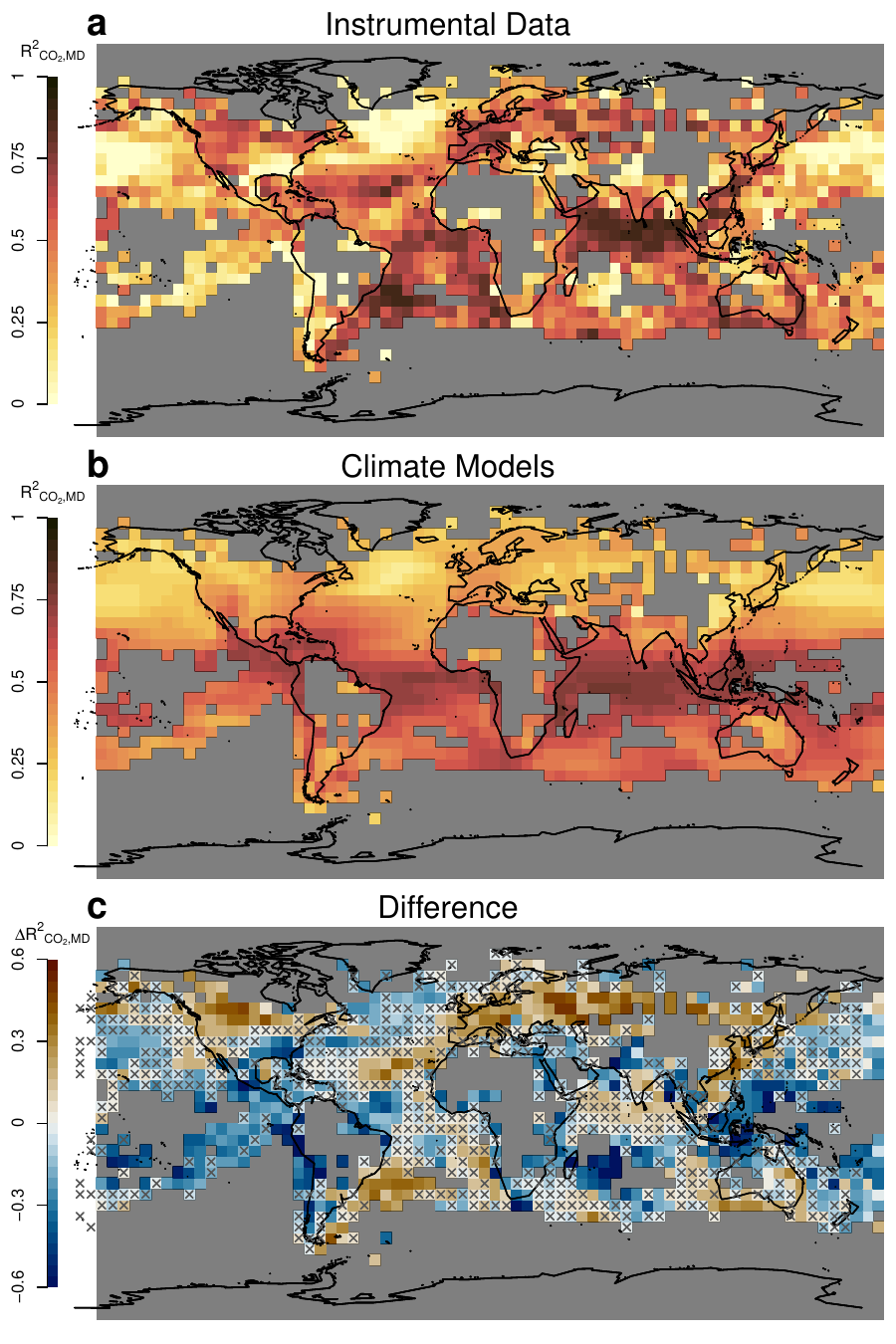}
\caption{Fraction of the multidecadal variance explained by the CO\textsubscript{2}-congruent variability (${R^2}_{CO_2,MD}$). \textbf{a}, Shown is the ensemble mean ${R^2}_{CO_2,MD}$ of the instrumental data. \textbf{b}, Same as \textbf{a}, but for the ensemble mean of the models. \textbf{c}, Difference between the instrumental (\textbf{a}) and model (\textbf{b}) estimates. The gray "$\times$" markers indicate that less than three quarter of the models agree on the sign of the difference.}\label{Fig_1}
\end{figure}

To avoid these pitfalls, we take a more conservative approach by removing only the forced variability in which we have the highest confidence—specifically the CO\textsubscript{2}-congruent variability. By locally subtracting the CO\textsubscript{2}-congruent variability through linear regression with the global CO\textsubscript{2} forcing, we derive the CO\textsubscript{2}-detrended temperature field $T_{DCO_2}(\overrightarrow{x},t)$\cite{wu_new_2019,hebert_regional_2018} (Methods). The latter retains natural (both forced and internal) variability and residual anthropogenic forced variability uncorrelated with the CO\textsubscript{2} forcing. In model experiments forced only by greenhouse gases, this CO\textsubscript{2}-detrending method yields results nearly identical to removing the single-model ensemble mean ($r=0.97 \pm 0.01$, Supp. Fig~\ref{fig:ghg}). The technique can be equally applied to instrumental data and historical model simulations, enabling a direct comparison between the residual forced, mostly related to anthropogenic aerosol forcing, and internal variability in the CO\textsubscript{2}-detrended temperature fields. To establish a baseline for comparison without forced variability, the internal variability from models $T_{Internal}(\overrightarrow{x},t)$ obtained by subtracting the single-model ensemble mean is also analyzed.
It is insightful to first quantify and study the multidecadal variance (hereafter, we use 30-80 years for calculations) explained by the CO\textsubscript{2}-congruent variability (${R^2}_{CO_2,MD}$; Methods). On average, it accounts for $41 \pm 2\%$ of the local multidecadal variance in the instrumental data, and \(46 \pm 14\%\) for the models (hereafter, the ensemble results are reported as mean \(\pm\) one standard deviation; note that the instrumental and model spread are distinct, corresponding to observational uncertainty in the former, and a combination of model structural uncertainty and internal variability in the latter). The mean spatial patterns of ${R^2}_{CO_2,MD}$ are broadly similar between the instrumental data and models over the ocean (\(r = 0.54\); \(p < 0.01\), for individual realizations \(r = 0.40 \pm 0.08\)) but differ over land (\(r = 0.1\); \(p > 0.1\), \(r = 0.08 \pm 0.08\) for individual realizations). The agreement over the oceans is largely driven by lower values of R²\textsubscript{CO\textsubscript{2},MD} in the North Pacific and North Atlantic, and higher values in equatorial regions of the Atlantic and Indian oceans (Fig.~\ref{Fig_1}). The lack of agreement over land relates to a larger share of multidecadal variance being explained by CO\textsubscript{2}-congruent variability in instrumental data over North America and Eurasia than expected from the models (Fig.~\ref{Fig_1}). This suggests a stronger direct response to the globally homogeneous CO\textsubscript{2} forcing in northern land regions\cite{hebert_regional_2018}, whereas in models higher levels of non-CO\textsubscript{2}-related variability, either from other forcings or internal processes, obscure the CO\textsubscript{2}-congruent variability. 

\begin{figure}[ht]

\includegraphics[width=0.5\linewidth]{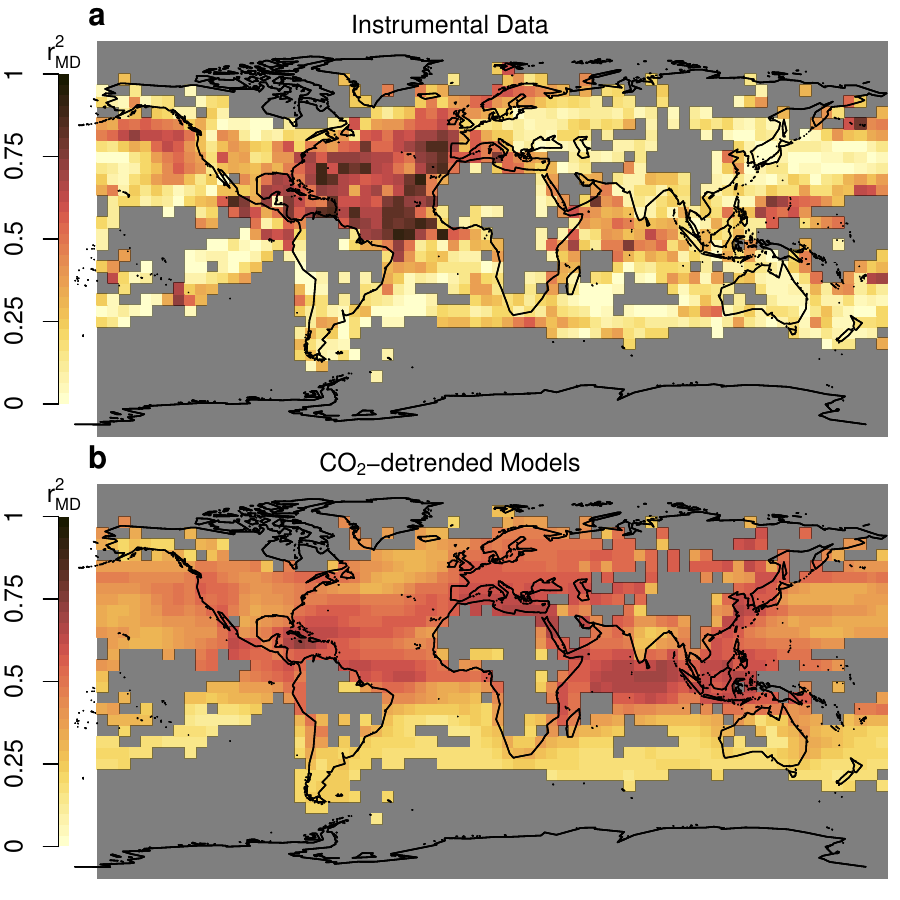}\includegraphics[width=0.5\linewidth]{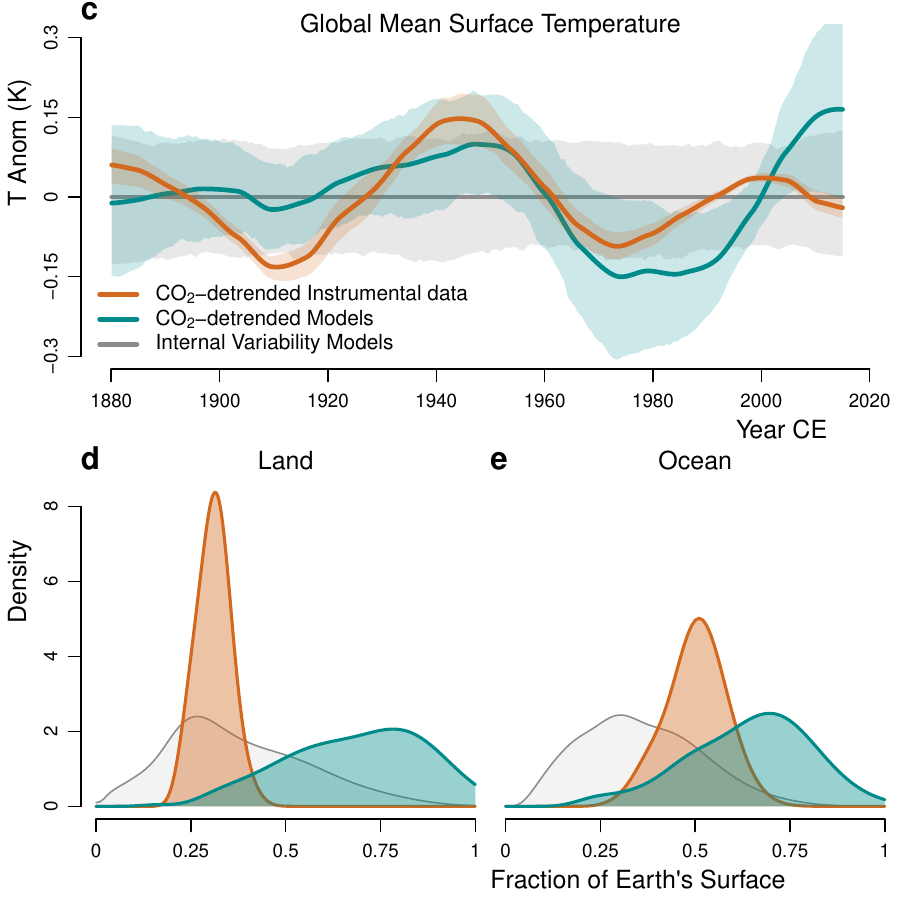}

\caption{Spatial extent of correlation patterns with GMST in instrumental and model data. \textbf{a}, Multidecadal variance explained by GMST in the local instrumental temperature timeseries after both were CO\textsubscript{2}-detrended. \textbf{b}, Same as \textbf{a}, but showing the average of the models. \textbf{c}, CO\textsubscript{2}-detrended GMST timeseries are shown for the instrumental data (orange) and the models (cyan); shading indicates 66\% spread of the respective ensembles. The analogous result for the internal variability only of the models is also shown (gray). \textbf{d}, Spatial extent of the patterns in a,b given as the fraction of the Earth's land area that was above the $r^2>0.25$ threshold; also shown for comparison is the same measure calculated from the internal variability in the models (gray distribution). \textbf{e}, Same as \textbf{d}, but for Earth's ocean area.}\label{Fig_2}
\end{figure}

To assess the relative importance of residual forced and internal temperature variability, we examine the spatial patterns of MDV associated with global mean surface temperature (GMST), as the two types of variability are expected to have distinct spatial signatures\cite{parsons_magnitudes_2020}. We computed the multidecadal squared correlation $r_{MD}^2$ between the CO\textsubscript{2}-detrended GMST and temperature field (Fig.~\ref{Fig_2}a,b,c; Methods). Although both instrumental and model data yield similar CO\textsubscript{2}-detrended GMST timeseries—capturing the early 20\textsuperscript{th} century warming, post-war cooling and post-1990 warming (Fig.~\ref{Fig_2}c)—the spatial patterns of squared correlation only exhibit similarity over the oceans again ($r=0.52$ for the models ensemble mean, $p<0.05$), but not over land ($r=0.06$ for the models ensemble mean, $p>0.1$, Fig.~\ref{Fig_2}a,b). This result is robust across individual model realizations, with greater agreement over the oceans ($r=0.30\pm0.12$) than over land ($r=0.11\pm0.10$). 

These correlation patterns highlight the different contributions of various regions to global mean MDV, offering insights in the historical variability of the CO\textsubscript{2}-detrended GMST timeseries. Specifically, they show that the North Atlantic, and to a lesser extent other oceanic regions, largely drive the global mean MDV in instrumental data, underscoring the dominant role of oceanic variability. In contrast, models exhibit more balanced contributions from both ocean and land regions in their global mean MDV.

A key diagnostic for identifying the origin of climate variability is the spatial covariance structure. Generally, forced variability, driven by large-scale or global factors such as stratospheric volcanic eruptions and aerosols, tends to produce spatially extensive fluctuations, whereas internal variability often results in more localized anomalies\cite{cleveland_stout_fingerprinting_2023}. One simple metric to assess this is the spatial extent of the correlation patterns (with GMST) above a given threshold; here, we use $r_{MD}^2>0.25$ (Methods). In the CO\textsubscript{2}-detrended instrumental data, the correlation pattern spans $45 \pm 6\%$ of the Earth's surface (Fig~\ref{Fig_2}a). In contrast, the CO\textsubscript{2}-detrended models show a more widespread coherence, covering $65\pm 15 \%$ of the Earth's surface, whereas their internal variability is indeed more localized, covering only $35 \pm 15\%$ of the Earth's surface. 

A more pronounced discrepancy emerges when considering the spatial extent of the correlation patterns over land and ocean separately. In the instrumental data, the correlation pattern is broader over the ocean ($50 \pm 7 \%$) than over land ($31 \pm 4 \%$), while in the models, they are of similar extent over both land ($69 \pm 17 \%$ for $T_{DCO_2}$ and $38 \pm 18 \%$ for $T_{Internal}$) and ocean  ($64 \pm 15 \%$ for $T_{DCO_2}$ and $36 \pm 15 \%$ for $T_{Internal}$; Fig.~\ref{Fig_2}d,e). The notable feature here is that in the instrumental data, the correlation pattern is significantly broader over the ocean than land by $19 \pm 6 \%$, whereas model simulations display nearly identical extents. This result is robust across models and methodological choices, and is supported by more general diagnostics such as the effective spatial degrees of freedom (Supp. Fig.~\ref{fig:models},~\ref{fig:esdof}). Thus, the main variations in the CO\textsubscript{2}-detrended global mean temperature are primarily driven by oceanic regions in the instrumental data, whereas land-sea boundaries do not appear to affect the extent of spatial anomalies in the models.

\subsection{Land-ocean contrast inversion}
The observed mismatch between instrumental and model data in terms of the spatial covariance structure of MDV is thus especially pronounced over land in the CO\textsubscript{2}-detrended temperature fields. This suggests different relative contributions of internal and (residual) forced variability on multidecadal timescales, as these components are expected to behave differently over land and ocean. The transient response to external forcing is known to be stronger over land compared to the ocean\cite{hebert_regional_2018} due to thermal inertia, as well as stronger shortwave cloud feedbacks\cite{sejas_feedback_2014}. In addition, moisture and evaporation feedbacks further amplify the temperature response over land as the change in equipotential temperature (moist energy) is equal; this also applies to internal oceanic-driven variability\cite{byrne_link_2013,sejas_feedback_2014,tyrrell_influence_2015}. When variability originating from the ocean transitions to land, the amplitude of the signal can also be damped depending on the strength of the ocean-land atmospheric coupling. Therefore, if MDV is mainly driven by external forcing\cite{booth_aerosols_2012,mann_absence_2020,mann_multidecadal_2021}, we expect larger variability on land than over the ocean, resulting in a land-ocean variance ratio above 1. However, if oceanic variability is the main driver\cite{kim_key_2018,li_investigation_2020,hebert_millennial-scale_2022}, it remains uncertain whether the ratio would be above or below 1 given the competing effects of the land-ocean coupling, and that of the moisture and evaporation feedbacks.

To illustrate the land-ocean contrast, we first consider two large oceanic and continental regions: the North Atlantic, and Central Asia, each spanning 60\textdegree in longitude and covering the mid-latitudes (25\textdegree~N-65\textdegree~N):  (Supp. Fig~\ref{fig:RLI}). In order to identify the common (or lack of) large-scale multidecadal climate signal across space, we take the regional mean surface temperature across the regions (see Supp. Fig.~\ref{Supp:CorImages} for a generalization across spatial scales). 

\begin{figure}[ht]

\includegraphics[width=.84\linewidth]{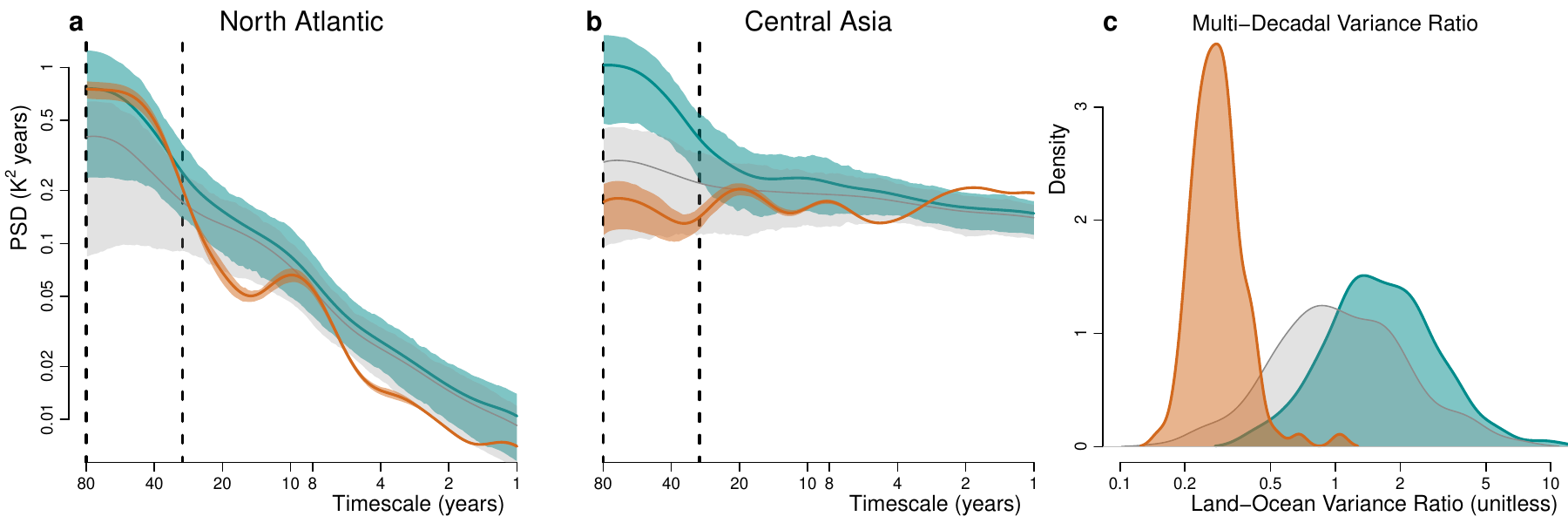}

\includegraphics[width=.9\linewidth]{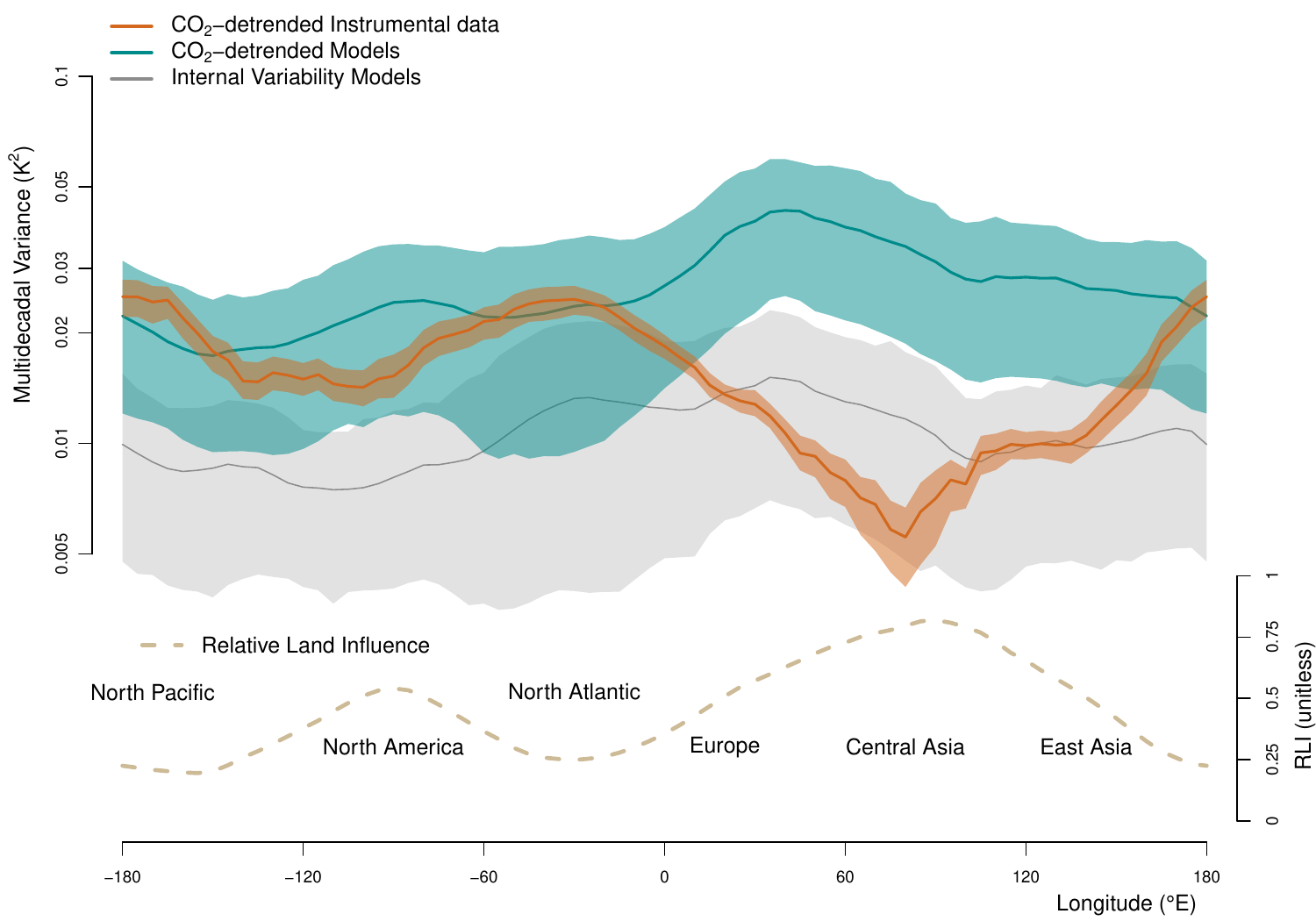}

\caption{Land-ocean constrast across the mid-latitudes. \textbf{a}, Power spectral density of CO\textsubscript{2}-detrended mean temperature for the North Atlantic region, comparing instrumental and model data. Shading indicates the 66\% spread of the respective ensembles, and the range of multidecadal timescales (30-80 years) is highlighted by vertical dashed lines. \textbf{b}, Same as \textbf{a}, but for the Central Asia region. \textbf{c}, Comparison of the land-ocean multidecadal variance ratio (Central Asia over North Atlantic) between CO\textsubscript{2}-detrended instrumental and model data. \textbf{d}, Opposite variations in multidecadal variance are observed between the CO\textsubscript{2}-detrended instrumental and model data across longitudinally sliding boxes over the northern mid-latitudes (see Supp. Fig.~\ref{ExtDataFig_3} for the undetrended result). Shading shows the 66\% spread of the respective ensembles.}\label{Fig_3}
\end{figure}

In the North Atlantic, both instrumental and model data show high MDV, regardless of whether they are CO\textsubscript{2}-detrended, indicating that a broadly coherent signal remains after spatial averaging (Fig.~\ref{Fig_3}a, Fig.~\ref{Fig_2}). In contrast, over Central Asia, while instrumental and model data initially exhibit similar MDV (Supp. Fig.~\ref{Fig:6Regions}d), they diverge once CO\textsubscript{2}-congruent variability is removed as it accounts for twice as much variance in the instrumental data than in the models (${R^2}_{CO_2,MD}=84 \pm 4 \%$ vs ${R^2}_{CO_2,MD}=44 \pm  22 \%$). Rather, the amplitude of the remaining MDV in the instrumental data closely matches the internal variability of the models, suggesting that the CO\textsubscript{2}-detrended MDV reflects predominantly internal dynamics (Fig.~\ref{Fig_3}b). 

The lower MDV in Central Asia from instrumental data is unlikely to stem solely from data quality issues, given that the large-scale CO\textsubscript{2}-congruent warming is well represented across regions. Instead, it points to increasing spatial heterogeneity with timescale over land (Fig.~\ref{Fig:6Regions}q). The variance ratio between the two regions is indicative of the land-ocean contrast, and, for the models, the ratio is above 1 when residual forced variability is present, but is more variable when there is only internal variability (Fig.~\ref{Fig_3}c). In the case of the instrumental data, the ratio exceeds 1 before removing the CO\textsubscript{2}-congruent variability and below 1 after, further indicating that minimal residual forced variability remains. 

We generalize the land-ocean contrast by calculating, within sliding 60\textdegree-wide longitudinal boxes across the mid-latitudes, the correlation $r_{\sigma_{MD}^2,RLI}$ between the multidecadal variance ($\sigma_{MD}^2$, Methods) and a measure of ocean-to-land atmospheric coupling, quantified using the relative land influence (RLI) metric\cite{mckinnon_spatial_2013} (Methods). The RLI measures the fraction of time air parcels spent over land versus the ocean before reaching the given location, with higher values indicating more continental regions (Supp. Fig.~\ref{fig:RLI}). A positive correlation between the temperature variability and the RLI implies a land-ocean variance ratio above 1, and a negative correlation implies the opposite (Supp. Fig.~\ref{Supp_Fig_LO_r}). 

When the timeseries are undetrended, we find positive correlations between multidecadal variance and RLI, again indicating a land-ocean variance ratio greater than 1 since forced variability dominates (Supp. Fig.~\ref{ExtDataFig_3}). This holds for both instrumental ($r_{\sigma_{MD}^2,RLI}=0.87$, $p<0.1$, assuming 4 DOFs given there are 6 independent 60\textdegree longitude boxes) and model data ($r_{\sigma_{MD}^2,RLI}=0.84$, $p<0.1$). When analyzing the CO\textsubscript{2}-detrended instrumental data, however, the relationship with RLI becomes negative ($r_{\sigma_{MD}^2,RLI}=-0.92$, $p<0.05$; Fig.~\ref{Fig_3}d), and contrasts with the analogous result from the models where a positive correlation remains ($r_{\sigma_{MD}^2,RLI}=0.78$, $p>0.1$, for the models mean), resulting in systematically opposite variations in multidecadal variance along the mid-latitudes (Fig.~\ref{Fig_3}d). 

The negative relationship in the instrumental data implies a land-ocean variance ratio below 1 and thus an inversion on multidecadal timescales in the land-ocean contrast favoring higher variance over oceans. This suggests a key role for ocean-to-land atmospheric coupling in damping MDV over land. The result is robust across various spatial smoothing scales and most pronounced in the northern mid-latitudes (Supp. Fig.~\ref{Supp:CorImages}). Therefore, while the models produce MDV comparable to instrumental data in the mid-latitudes outside of Eurasia, they may be spuriously compensating for missing oceanic-driven internal variability with forced variability. This leads to inflated MDV in the models over Eurasia, where instrumental data better aligns with the internal variability of the models (Fig.~\ref{Fig_3}d). 

\begin{figure}
    \centering
    \includegraphics[width=0.666\textwidth]{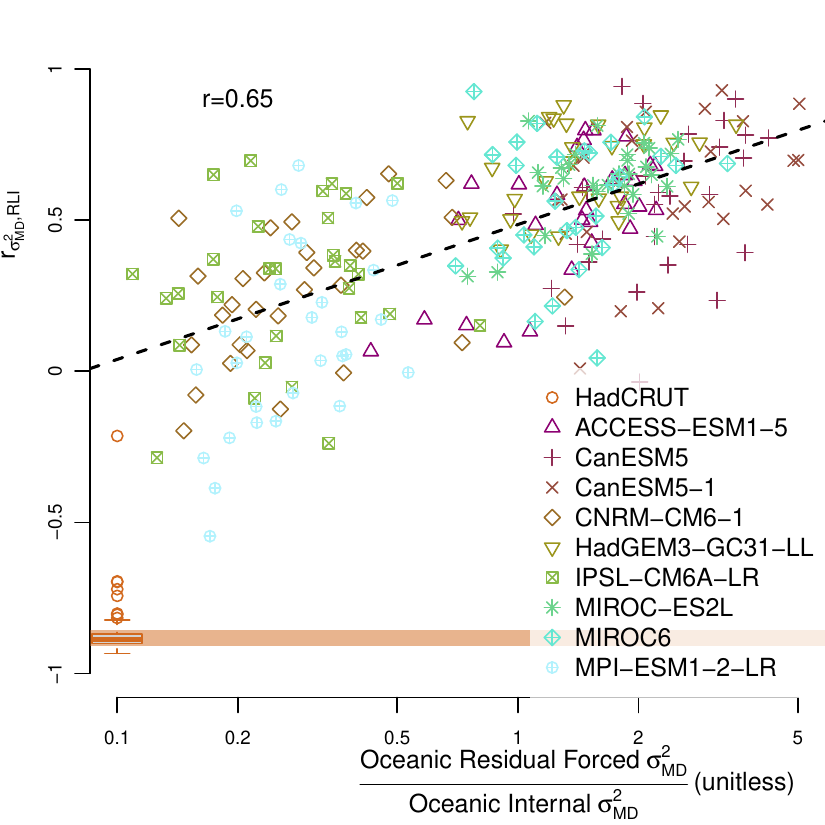}
        \caption{Relationship between the land-ocean contrast in MDV and the relative contribution of internal vs forced oceanic variability. The correlation  $r_{\sigma_{MD}^2,RLI}$ calculated from the CO\textsubscript{2}-detrended models across the northern mid-latitudes is shown as a function of the logarithm of the ratio of residual forced to internal oceanic multidecadal variance. Residual forced variability is estimated from the CO\textsubscript{2}-detrended single-model ensemble means, and internal variability from the individual realizations after the single-model ensemble means were removed. The regression line (dashed black) and corresponding correlation value are shown, highlighting the direct relationship between the driver of oceanic variability and the land-ocean contrast strength. The   $r_{\sigma_{MD}^2,RLI}$ values calculate for the instrumental data are shown as a boxplot for comparison (orange boxplot, circles indicate outliers) and arbitrarily placed at the leftmost value on the x-axis; the 66\% range is shown across with orange shading.}
    \label{Fig_4}
\end{figure}

The measured strength of the land-ocean contrast in the mid-latitudes, quantified here using $r_{\sigma_{MD}^2,RLI}$, may thus provide a useful constraint for evaluating climate models in terms of the relative importance of forced and internal MDV in oceanic regions. For each model run, we estimate the contribution of each variability type to oceanic MDV (taken as the average multidecadal variance of the North Pacific and North Atlantic mean temperatures) and examine its relationship with the strength of the land-ocean contrast (Fig.~\ref{Fig_4}). 

We find a significant direct relationship between the amplitude of the residual forced variability (calculated from the CO\textsubscript{2}-detrended single-model ensemble means) in the oceanic regions and stronger land-ocean contrast ($r=0.83$, $p<0.01$, $n=9$; Supp. Fig.~\ref{Supp_Fig_4}a), whereas the converse is true for internal variability ($r=-0.46$, $p<0.001$, $n=270$; Supp. Fig.~\ref{Supp_Fig_4}b). In fact, the ratio of residual forced to internal oceanic MDV appears to strongly determine the strength and direction of the land-ocean contrast in the mid-latitudes ($r=0.65$, $p<0.001$, $n=270$; Fig.~\ref{Fig_4}c). This shows that the origin of the oceanic variability matters as forced variability fosters a positive land-ocean contrast and internal variability a negative one, resulting in a robust emergent relationship to evaluate individual models.

Notably, the three models with the most often negative correlation values between variability and relative land influence, MPI-ESM1-2-LR, CNRM-CM6-1 and IPSL-CM6A-LR, also have the three weakest oceanic residual forced MDV (Supp. Fig.~\ref{Supp_Fig_4}a). In addition, MPI-ESM1-2-LR and IPSL-CM6A-LR are, among the models we considered, the two with the weakest present-day aerosol forcing (see Table 1 in Flynn et al., 2023\cite{flynn_strong_2023}). The three models also have, on average, the highest oceanic internal MDV, alongside HadGEM3-GC31-LL (Supp. Fig.~\ref{Supp_Fig_4}b). The latter, however, has the most positive  $r_{\sigma_{MD}^2,RLI}$ values as a result of also having the highest oceanic residual forced MDV (Fig.~\ref{Fig_4}a, Supp. Fig.~\ref{Supp_Fig_4}a), largely attributable to its high present-day aerosol forcing\cite{flynn_strong_2023}. 

While it is clear that a ratio in favor of internal, rather than forced, variability on multidecadal timescales reduces the land-ocean contrast in the mid-latitudes, even the best performing models do not consistently reproduce the inversion, towards higher variability over the ocean than land, observed in the instrumental data. This may result from a combination of more variable as well as more localized atmospheric circulation patterns than currently simulated by models, aligning with their known underestimation of the frequency of blocking events\cite{davini_cmip3_2020}. These findings further support the hypothesis that temperature fields should remain more spatially heterogeneous on longer timescales and that models tend to become overly coherent\cite{laepple_regional_2023}. It is plausible that models allow oceanic influence to reach more continental areas too consistently, or, in other words, that the relative land influence gradient going inland is too weak. Evaluating the ability of climate models to accurately reproduce realistic relative land influence fields could explain why no model consistently captures the observed inversion across temporal scales in the land-ocean contrast.

\subsection{Conclusion}
Our analysis suggests a fundamental mismatch between instrumental data and climate models in representing the spatial and temporal patterns of multidecadal temperature variability (MDV), particularly across the mid-latitudes. 

Specifically, CO\textsubscript{2}-congruent variability constitutes a larger fraction of the multidecadal variance in instrumental data, particularly over land, whereas spatially coherent residual forced variability with no apparent land-sea boundaries remains in models after CO\textsubscript{2}-detrending (Fig.~\ref{Fig_1},~\ref{Fig_2}).

In the last 150 years, instrumental data suggest a clear dominance of oceanic regions,  particularly the North Atlantic, in driving global mean temperature variability, supporting an oceanic origin to the MDV (Fig.~\ref{Fig_2}). The pronounced oceanic variability observed in instrumental records results in an inversion in the land-ocean contrast: at sub-decadal timescales, variability is higher over land, whereas at multidecadal timescales, it is higher over oceans (Fig.~\ref{Fig_3}) . This finding, based only on instrumental data, supports previous conclusions derived from pollen-based temperature reconstructions over the Holocene, which identified oceanic influence as the main driver of low-frequency temperature variability\cite{hebert_millennial-scale_2022},

We identified an emergent relationship in the climate model ensembles between the origin of the oceanic variability, whether forced or internal, and the land-ocean contrast (Fig.~\ref{Fig_4}). This relationship, along with the instrumental land-ocean contrast, suggests that most climate models lack sufficient low-frequency internal variability, consistent with findings based on paleoclimate records\cite{laepple_ocean_2014,hebert_millennial-scale_2022,cheung_comparison_2017}. Consequently, climate models may artificially compensate for this deficiency with volcanic and aerosol forcing in order to simulate multidecadal variability with the right amplitude and in phase with observations\cite{booth_aerosols_2012,chylek_atlantic_2014}. The forcing concurrently leads to an amplified and spatially coherent forced response over mid-latitude land areas and creates a pronounced land-ocean contrast, providing further evidence that strong aerosol forcing is incompatible with observations\cite{stevens_rethinking_2015,albright_origins_2021}. A less negative aerosol forcing would imply smaller transient climate sensitivity estimates and thus reduced warming projections in the near future\cite{hebert_regional_2018}, particularly over land regions, as the aerosol load is expected to be reduced\cite{watson-parris_large_2022}.

Meanwhile, models with higher internal variability have a weaker and sometimes inverted land-ocean contrast aligning better with instrumental data. This adds support to the existence of low-frequency modes of oceanic variability as observed in many climate models\cite{delworth_interdecadal_1993,wills_ocean_2019,latif_natural_2022}, but not all\cite{mann_absence_2020}. However, the question of whether such modes are robust cannot be fully resolved through modeling alone, as oscillatory behavior in models are emergent properties arising from many factors such as tuning strategies, parameterization choices, or resolution changes.

Advances in paleoclimate reconstructions are thus crucial to identify and characterize low-frequency modes of oceanic variability and inform climate model development. While there is evidence of significant Atlantic\cite{michel_early_2022} and Pacific MDV modes\cite{bruun_heartbeat_2017}, they mostly rely on terrestrial records, such as tree rings that are affected by low-frequency noise\cite{mcpartland_separating_2024}. Future advances in high-resolution marine proxy development\cite{obreht_last_2022}, proxy-system modeling and data assimilation techniques may help overcome these obstacles by harmonizing land and ocean archives, offering a more robust framework for evaluating low-frequency variability.

Our conclusions underscore the need to improve the representation of oceanic internal variability as regions with strong oceanic influence, where a large portion of the global population lives, may be subject to persistent large-scale climate changes that fall outside the envelope of currently simulated natural variability.

\subsection{Acknowledgements}
This is a contribution to the SPACE ERC project and was supported by AWI's INSPIRES program. The work profited from discussions at the CVAS working group of the Past Global Changes (PAGES) programme. We thank A. Dolman, and V. Skiba for discussion and feedback throughout the course of this study, and M. Casado, P. Karami an R. Zhang for providing written comments. We gratefully acknowledge the Earth System Grid Federation (ESGF) for providing access to the model data used in this study, and for their continued contribution to open climate data sharing. The authors declare no competing interests.

\FloatBarrier
\pagebreak
\bibliographystyle{unsrt}
\bibliography{LowfreqModels}
\pagebreak

\section{Supplementary Materials}
\setcounter{figure}{0}
\renewcommand{\tablename}{Supplementary Table}
\renewcommand{\figurename}{Supplementary Figure}
\renewcommand{\thepage}{S\arabic{page}}
\renewcommand{\thetable}{S\arabic{table}}
\renewcommand{\thefigure}{S\arabic{figure}}

\subsection{Materials and Methods}
\subsubsection{Data}
We utilize global and regional surface temperature data from instrumental observations as well as historical climate model simulations. The instrumental data consists of 100 members of the HadCRUT5 ensemble, spanning from 1850 to the present\cite{morice_updated_2021}, albeit we do not use the pre-1880 period given the sparseness of observations. The multi-model ensemble consists of CMIP6 historical simulations, from 9 models with 30 realizations each, spanning from 1850 to 2015. See Supplementary Table~\ref{tab:cmip6_models} for details. 

\subsubsection{Power Spectral Density}
Gaps in the climate model fields were introduced to match the data availability in the HadCRUT5 dataset\cite{morice_updated_2021}. Linear interpolation was applied at locations with at least 70\% data coverage from 1880 to 2015. The power spectral density was then calculated using the multitaper method with 3 tapers and a time-bandwitdh parameter of 2. The spectra were smoothed using a Gaussian kernel with a standard deviation of 0.1 in the logarithm (base 10) of timescale/frequency. Variance over a given timescale band was calculated by integrating the power spectral density over the corresponding frequency band.  

\subsubsection{CO\textsubscript{2}-Congruent Variability}
To account for the first-order anthropogenic forced response, we use the following temperature decomposition\cite{hebert_regional_2018}:
\begin{equation}
T(\overrightarrow{x},t) \approx S_{CO_2}(\overrightarrow{x}) \log{\rho_{CO_2}(t)}+T_{DCO_2}(\overrightarrow{x},t)
\end{equation}
where $S_{CO_2}(\overrightarrow{x})$ represents the spatially varying local transient climate sensitivity, $\rho_{CO_2}(t)$ is the global atmospheric CO\textsubscript{2} concentration\cite{meinshausen_rcp_2011}, and $T_{DCO_2}(\overrightarrow{x},t)$ is the residual temperature variability. We estimate $S_{CO_2}(\overrightarrow{x})$ locally through linear regression with $\log \rho_{CO_2}(t)$. The resulting temperature field  $T_{DCO_2}(\overrightarrow{x},t)$ contains residual forced variability uncorrelated with CO\textsubscript{2} forcing, as well as internal variability.

To quantify CO\textsubscript{2}-congruent multidecadal variance, we compare the power spectra of the temperature fields before and after removing the CO\textsubscript{2}-congruent variability: 
\begin{equation}
{R^2}_{MD,CO_2}(\overrightarrow{x})=1-\frac{{\sigma^2}_{MD,DCO_2}(\overrightarrow{x})}{{\sigma^2}_{MD}(\overrightarrow{x})}
\end{equation}
where $\sigma^2_{MD}(\overrightarrow{x})$ and $\sigma^2_{MD,DCO_2}(\overrightarrow{x})$ represent the mean power spectra density over the 30-80 year timescale band for the undetrended and CO\textsubscript{2}-detrended residual temperature fields, respectively.

\subsubsection{Temporal Smoothing}
To isolate multidecadal variance in timeseries, we apply the locally estimated scatterplot smoothing (LOESS) method with a 15-year span, effectively producing results comparable to a 30-year lowpass filter. The multidecadal correlation $r_{MD}$ between the GMST and local temperature timeseries (Fig.~\ref{Fig_2}) is calculated after applying a 15-year LOESS. In order to preserve the full length of the timeseries, padding is applied under the minimum slope constraint by reflecting the data around the boundary points\cite{mann_smoothing_2004}.

\subsubsection{Threshold for Correlation Patterns}
We set a threshold of ${r_{MD}}^2=0.25$ to measure the spatial extent of multidecadal correlation patterns (Fig.~\ref{Fig_2}). The threshold represents the 90\textsuperscript{th} percentile of the distribution of squared correlation expected from 135-year long timeseries of white noise, smoothed with the 15-year LOESS filter. Importantly, the difference in spatial extent of correlation patterns between instrumental and model data is robust to the choice of threshold. For example, similar conclusions hold for a threshold of ${r_{MD}}^2=0.43$, which can be derived using the same methodology for 1/f noise instead of white noise.

\subsubsection{Effective Spatial Degrees of Freedom}
To determine the effective spatial degrees of freedom (ESDOF) for a region or the entire Earth, we calculate two metrics and take their ratios. (i) Mean Local Power Spectrum: We first compute the power spectra at all locations and average them in the spectral domain, resulting in the mean local power spectrum. (ii) Power Spectrum of the Global (or Regional) Mean: Next, we compute the average timeseries across all locations (i.e. in the temporal domain), and compute its power spectrum, yielding the power spectrum of the global (or regional) mean. The ESDOF is then defined as the frequency-dependent ratio of the mean local power spectrum (i) to the power spectrum of the global (or regional) mean (ii)\cite{kunz_frequency-dependent_2021}. Since the power spectrum of the mean is based on a single timeseries, the spectral estimates are more variable. To stabilize the ESDOF estimates, we apply smoothing to both power spectra before taking the ESDOF ratio. The smoothing is performed using a Gaussian kernel with a standard deviation of 0.1 in the logarithm (base 10) of timescale/frequency.

\subsubsection{Relative Land Influence}
The Relative Land Influence (RLI) metric quantifies the extent to which air parcels in the atmospheric column have previously traveled over land or ocean. Using the HYbrid Single-Particle Lagrangian Integrated Trajectory (HYSPLIT) parcel trajectories, RLI was computed by McKinnon et al. (2013)\cite{mckinnon_spatial_2013} as the weighted average of land and ocean surface influences based on whether air parcels were over land or ocean at each 3-hour interval. RLI values reflect a spatial gradient, with increasing land influence from west to east across continents in the northern hemisphere, and vice versa over ocean basins.  This metric provides insight into the land-ocean atmospheric connectivity and has been applied to understand regional climate variability\cite{hebert_millennial-scale_2022}.

For detailed computation, refer to McKinnon et al. (2013)\cite{mckinnon_spatial_2013}.
\FloatBarrier
\subsection{Supplementary Table}
\begin{table}[ht]
\centering
\begin{tabular}{l p{5 cm} l}
\hline
\textbf{Model} & \textbf{Institution} & \textbf{Publication} \\ \hline
ACCESS-ESM1-5  & Commonwealth Scientific and Industrial Research Organisation & Ziehn et al. (2020)\cite{ziehn_australian_2020}\\ 
CanESM5        & Canadian Centre for Climate Modelling and Analysis & Swart et al. (2019)\cite{swart_canadian_2019} \\
CanESM5-1      & Canadian Centre for Climate Modelling and Analysis & Sigmond et al. (2023)\cite{sigmond_improvements_2023}\\
 CNRM-CM6-1& Centre National de Recherches Météorologiques&Voldoire et al.(2019)\cite{voldoire_evaluation_2019}\\ 
HadGEM3-GC31-LL & Met Office Hadley Centre & Williams et al. (2018)\cite{andrews_historical_2020}\\ 
IPSL-CM6A-LR   & Institut Pierre-Simon Laplace & Boucher et al. (2020)\cite{boucher_presentation_2020}\\ 
MIROC-ES2L     & Japan Agency for Marine-Earth Science and Technology & Hajiba et al. (2020)\cite{hajima_development_2020} \\ 
MIROC6         & Japan Agency for Marine-Earth Science and Technology & Tatebe et al. (2019)\cite{tatebe_description_2019}\\ 
MPI-ESM1-2-LR  & Max Planck Institute for Meteorology & Mauritsen et al. (2019)\cite{mauritsen_developments_2019}\\ \hline
\end{tabular}
\caption{CMIP6 models used in this study. For each model, 30 full-forcing historical experiments were available. In the case of HadGEM3-GC21-LL, MIROC6 and MPI-ESM1-2-LR, 30 greehouse gas-only historical experiments were additionally available.}
\label{tab:cmip6_models}
\end{table}
\FloatBarrier
\subsection{Supplementary Figures}
\begin{figure}
    \centering
    \includegraphics[width=\textwidth]{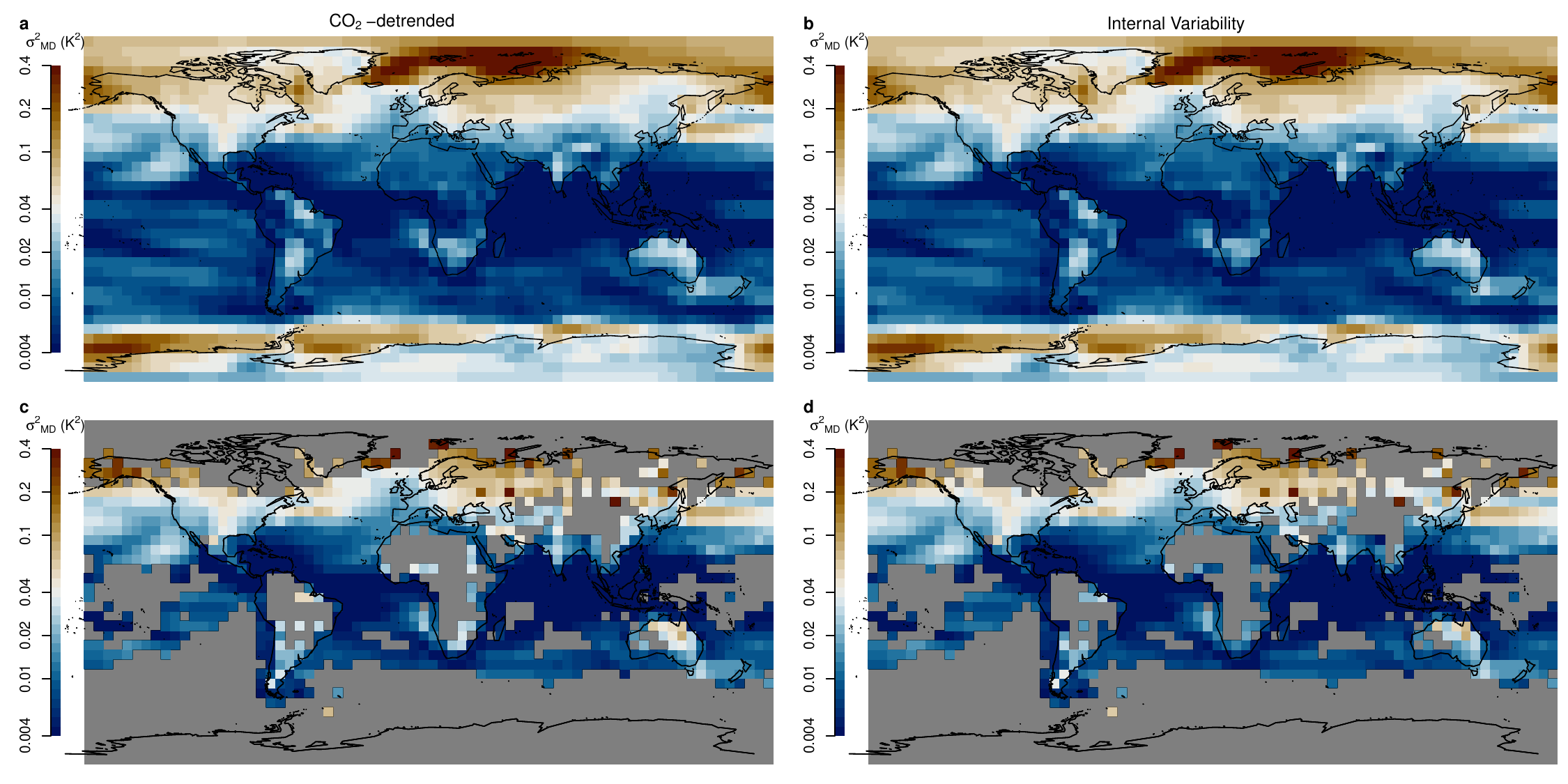}
\caption{Evaluation of CO\textsubscript{2}-detrending in greenhouse gas only (GHG-Only) historical experiments versus subtracting the single-model ensemble mean. \textbf{a}, Multidecadal variance (30-80 years) calculated from the power spectra of the CO\textsubscript{2}-detrended temperature fields, averaged across the GHG-only ensemble (comprising 90 realizations across three models: HadGEM3-GC31-LL, MIROC6 and MPI-ESM1-2-LR). \textbf{b}, Same as a, but with internal variability obtained by subtracting the single-model ensemble mean from each realisation. \textbf{c}, Same as a, but introducing data gaps matching the HadCRUT instrumental data sampling before the CO\textsubscript{2}-detrending is performed. \textbf{d}, Same as b, but also with sampling gaps as in c. The spatial patterns of multidecadal variance obtained by both methods are highly similar and strongly correlated ($r=0.99 \pm 0.01$ without sampling gaps, $0.97 \pm 0.01$ with sampling gaps). }
    \label{fig:ghg}
\end{figure}

\begin{figure}
    \centering
    \includegraphics[width=\textwidth]{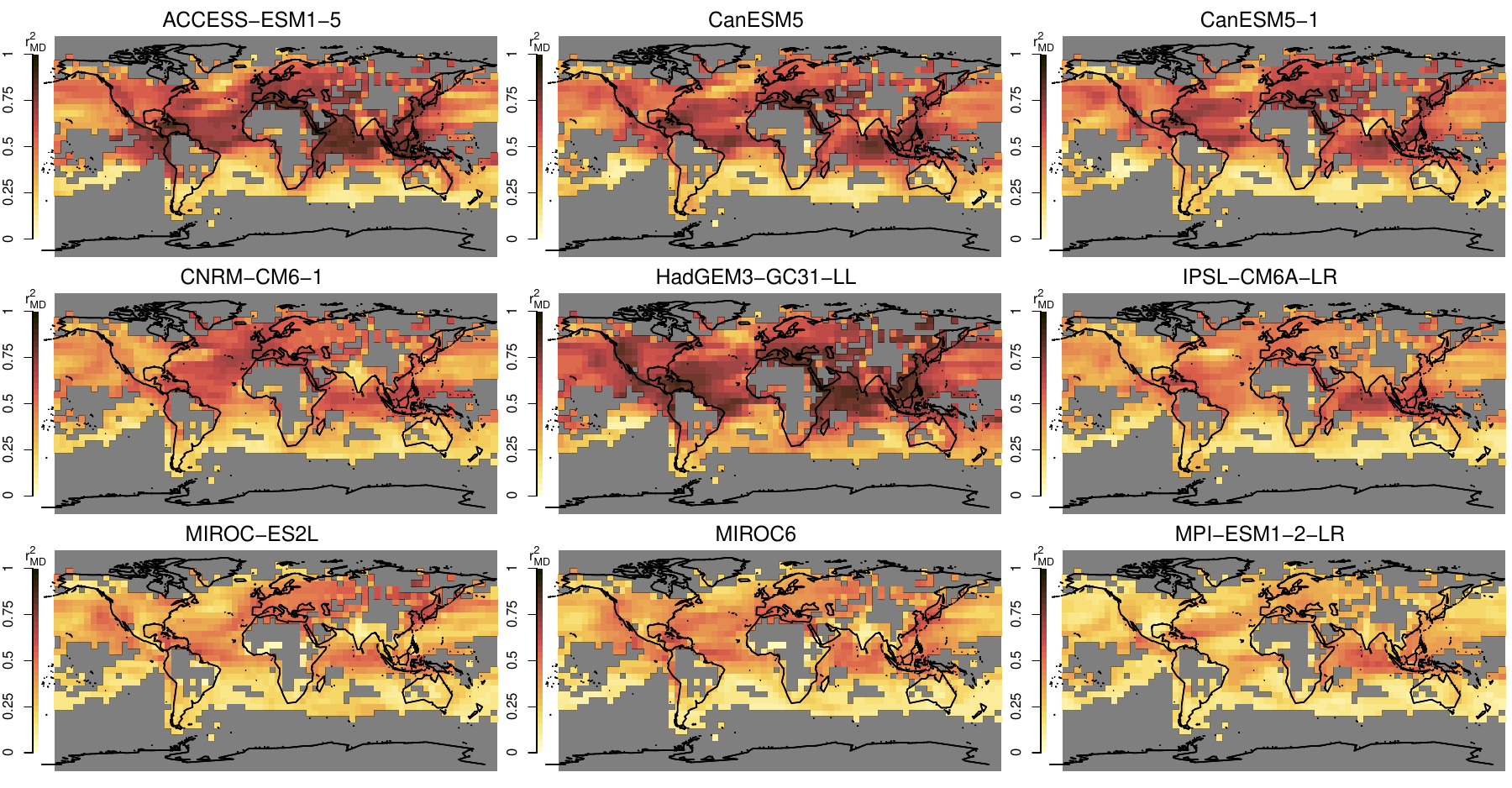}
\caption{Spatial extent of correlation patterns with GMST in CMIP6 models after they were CO\textsubscript{2}-detrended; same as in Fig.~\ref{Fig_2}b, but showing the individual models separately. For each model, the average of 30 realizations is shown.}
    \label{fig:models}
\end{figure}

\begin{figure}
    \centering
    \includegraphics[width=\textwidth]{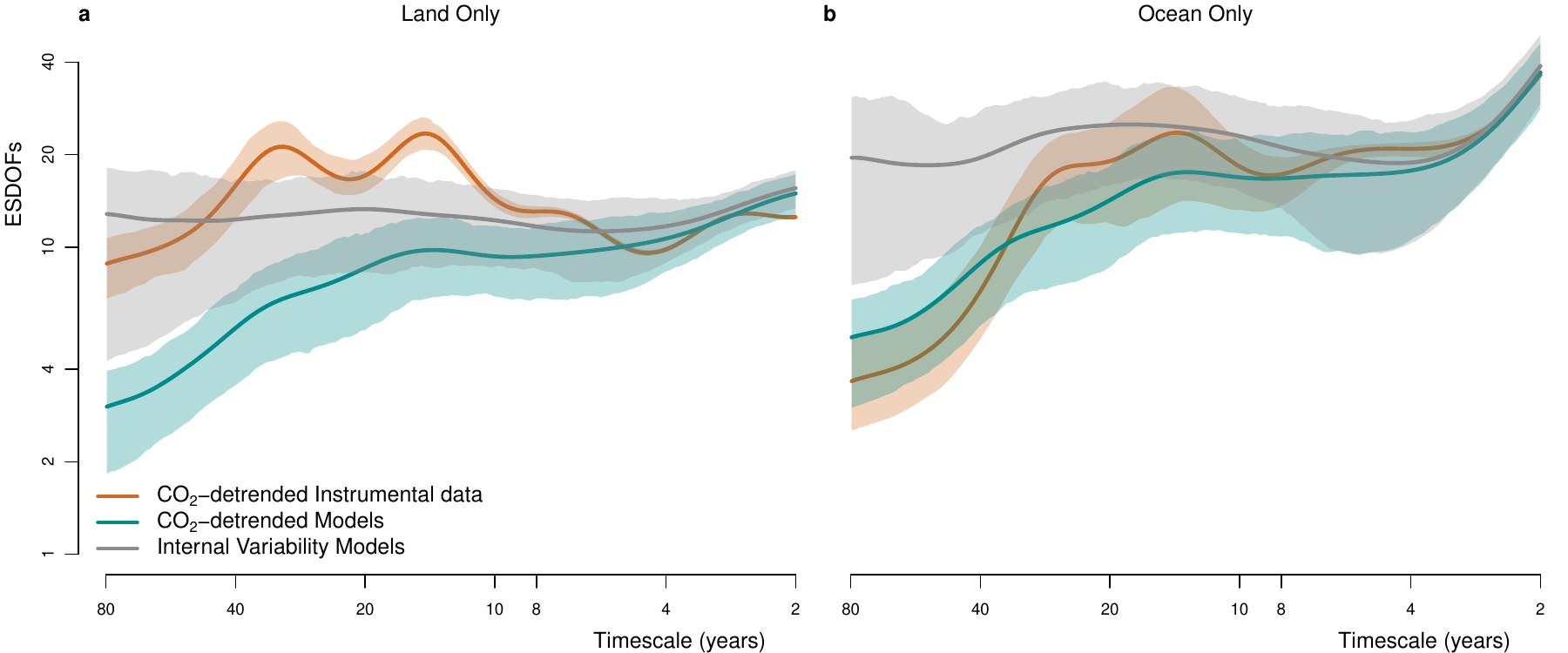}
    \includegraphics[width=\textwidth]{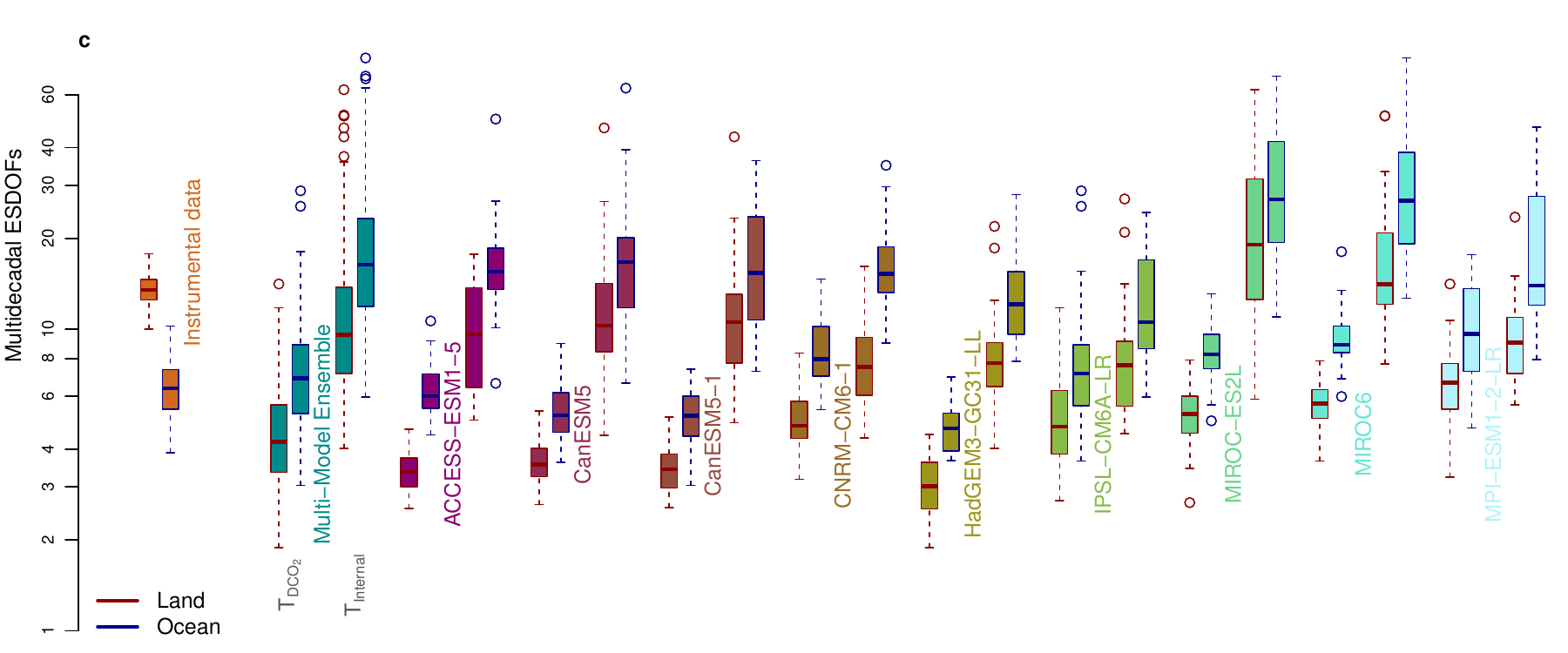}
    \caption{Effective Spatial Degrees of Freedom (ESDOFs) across land and ocean regions. \textbf{a}, ESDOF for land areas, comparing CO\textsubscript{2}-detrended instrumental data (orange), model data (cyan), and internal variability from the models (gray). \textbf{b}, Same as a, but for the oceans. \textbf{c}, Boxplots showing the range of multidecadal (30-80 year) ESDOFS for the instrumental data (n=100), multi-model ensemble (n=270) and individual single model ensembles (n=30 each), estimated separately for land (red border) and ocean (blue border). For each ensemble/model, the boxplots display the results for the CO\textsubscript{2}-detrended data (left of the name) and internal variability (right of the name). The latter case was not available for the instrumental data.  Model/ensemble names are color-coded to match the associated boxplot fills.}
    \label{fig:esdof}
\end{figure}

\begin{figure}
    \centering
    \includegraphics[width=\textwidth]{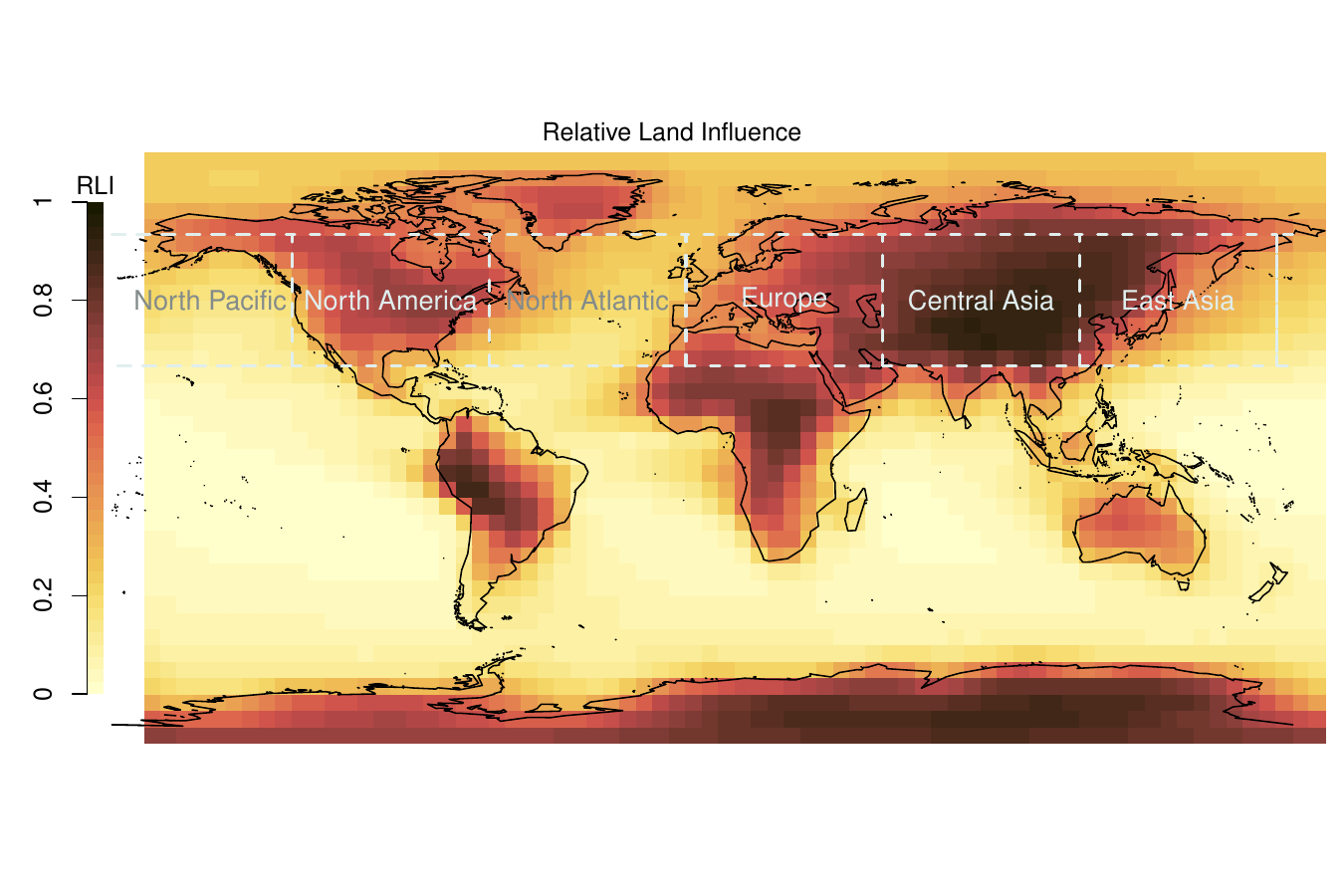}
    \caption{Relative Land Influence (RLI)\cite{mckinnon_spatial_2013}, representing the fraction of time that air parcels traveled over land vs the ocean before reaching the given location. Shown are six example regions used in the analysis. The average RLI for the six regions are, respectively from west to east on the map, 0.21,0.54,0.27, 0.56, 0.84 and 0.50.} \label{fig:RLI}
\end{figure}
    
\begin{figure}
    \centering
    \includegraphics[width=0.75\textwidth]{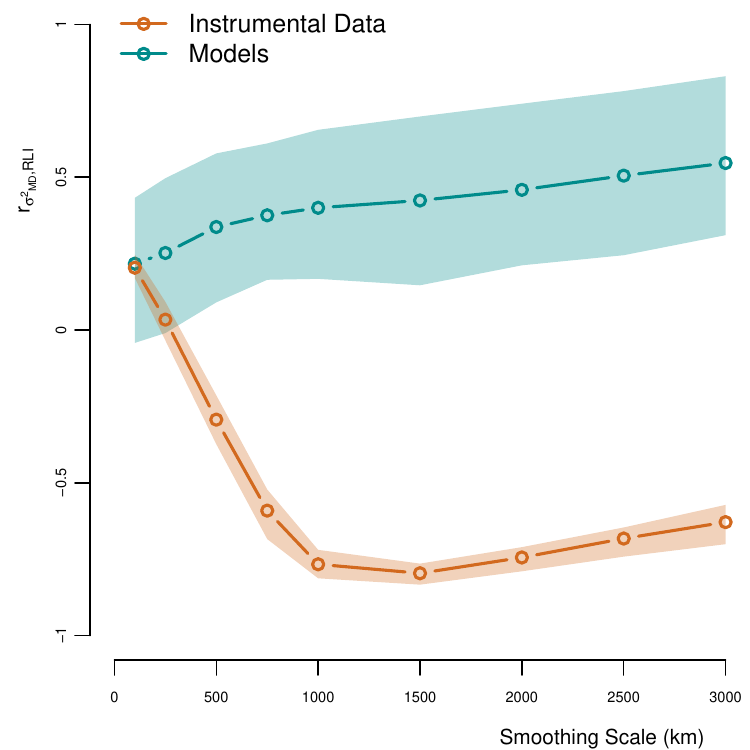}
    \caption{Generalization of the land-contrast analysis for different smoothing lengths and latitudes. The correlation  $r_{{\sigma^2}_{MD},RLI}$ between the multidecadal variance ${\sigma^2}_{MD}$ and the relative land influence (RLI)  for the instrumental and model data for different (Gaussian) smoothing lengths over the mid-latitudes. A negative  $r_{{\sigma^2}_{MD},RLI}$ is found in the northern mid-latitudes from the instrumental data for smoothing lengths longer than 250 km, while the average  $r_{{\sigma^2}_{MD},RLI}$ from the models is positive and increases with smoothing length. The main analysis is performed with 60\textdegree-wide in longitude rectangular boxes, which have a spatial extent similar to a 2000 km Gaussian smoothing.}\label{Supp:CorImages}
\end{figure}

    \begin{figure}[ht]

\includegraphics[width=.9\linewidth]{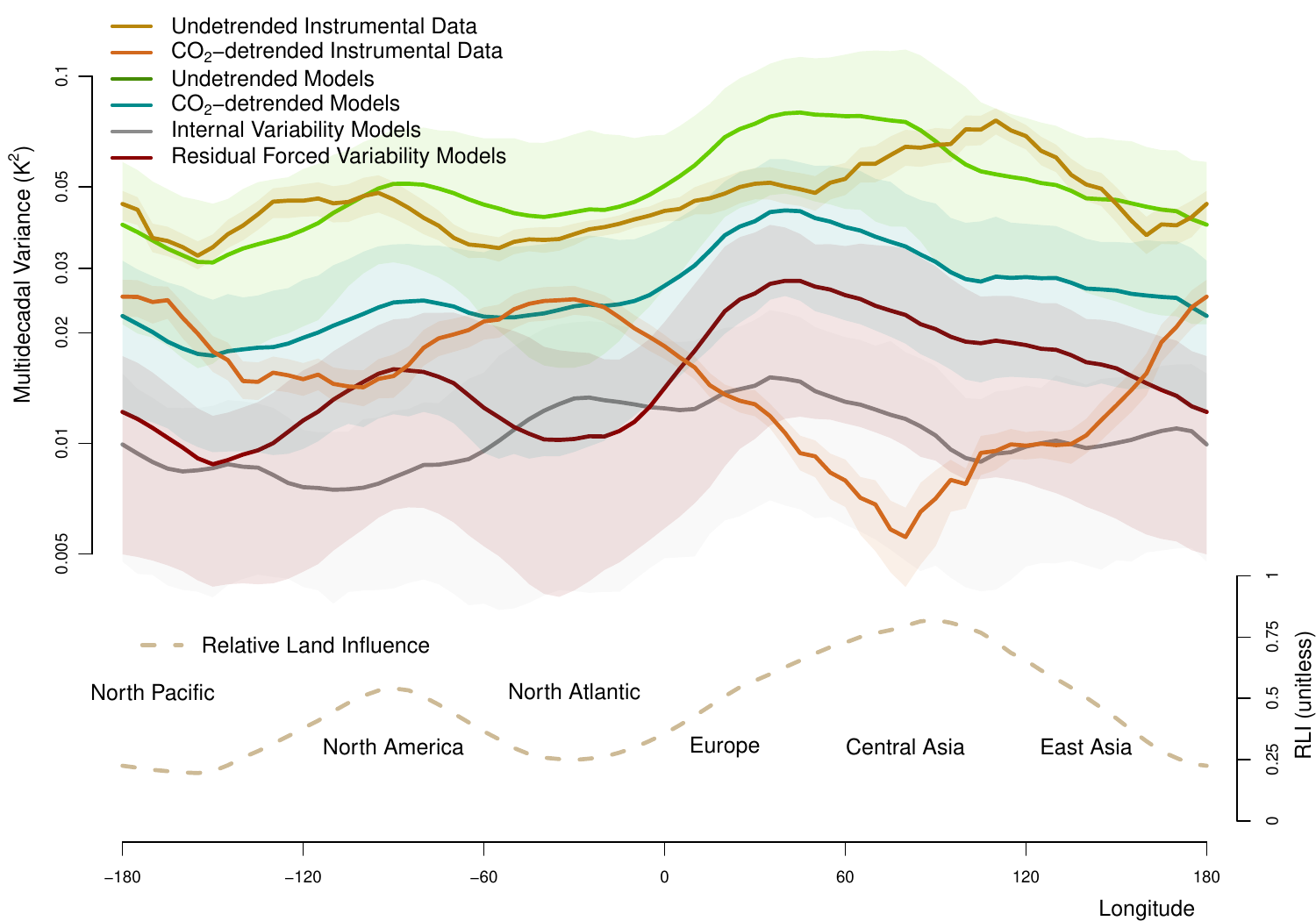}

\caption{Same as Fig.~\ref{Fig_3}, but also including the results for the undetrended datasets, thus dominated by anthropogenic forced variability (gold for instrumetnal data, and green for models). The Relative Land Influence (RLI)\cite{mckinnon_spatial_2013} shows a strong positive correlation with the mutli-decadal variance for both the instrumental data ($r=0.87$, $p<0.1$), and the models ($r=0.84$,$p<0.1$). In addition, we also show the result for the CO\textsubscript{2}-detrended single model ensemble means, thus corresponding to the residual forced variability only ($r=0.87$, $p<0.1$, red), i.e. without the internal variability}\label{ExtDataFig_3}
\end{figure}

\begin{figure}
    \centering
    \includegraphics[width=\textwidth]{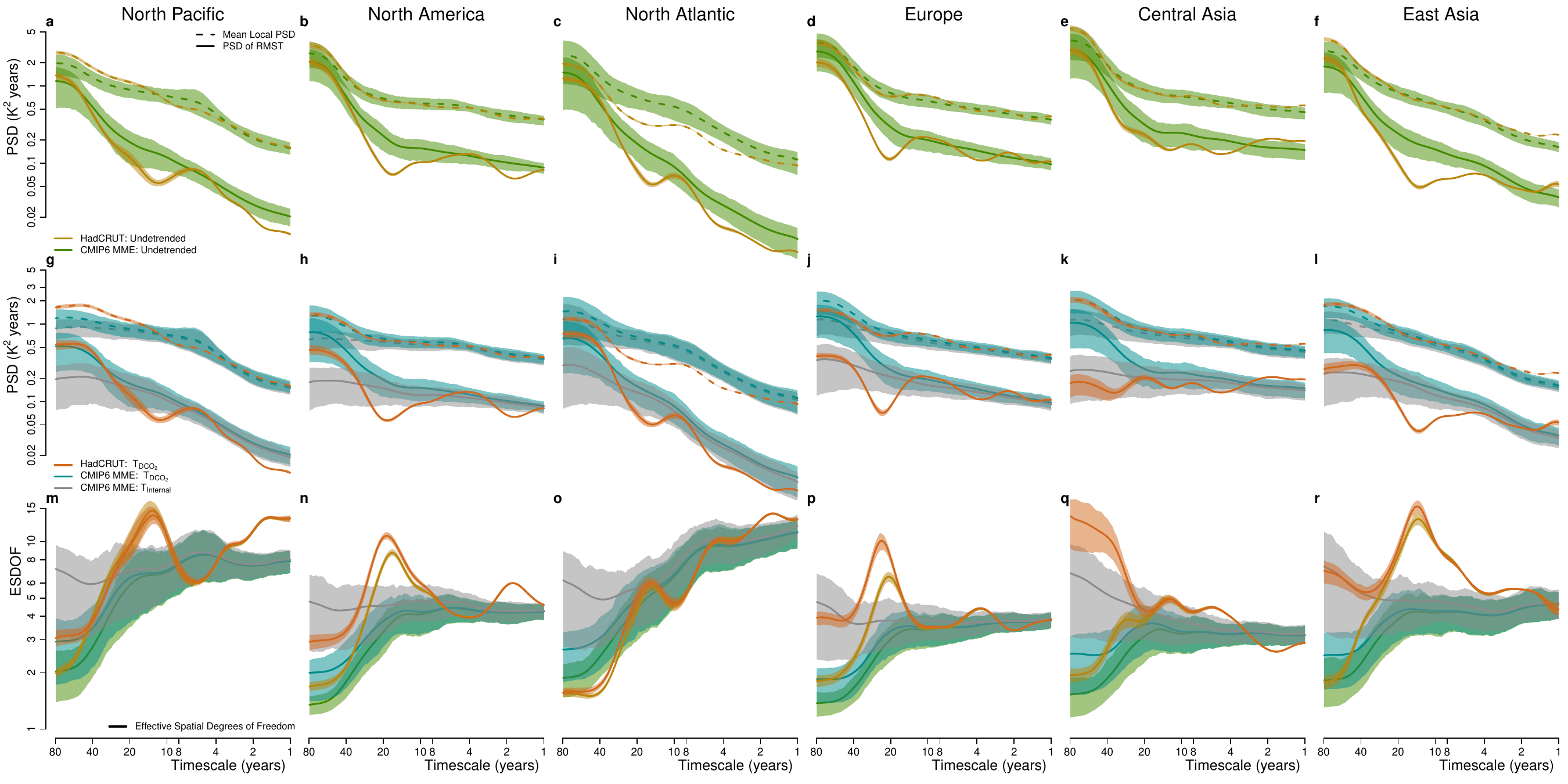}

    \caption{\textbf{a,b,c,d,e,f}, The mean local power spectral density (PSD; dashed lines) within the six mid-latitude regions (Supp. Fig.~\ref{fig:RLI}) is compared with the PSD of the undetrended regional mean surface temperature of the same regions (solid), for both the instrumental (gold) and model data (green). \textbf{g,h,i,j,k,l}, Same as a-f, but for the CO\textsubscript{2}-detrended data. Additionally, the PSD of the internal variability component of the models is included (gray). \textbf{m,n,o,p,q,r}, The ratio of the mean local PSD (dashed lines in a-l) to the PSD of the regional mean temperature (solid lines in a-l) provide an estimate of the Effective Spatial Degrees of Freedom (ESDOFs) for the respective temperature fields (Methods). Generally, when the data are not detrended, both instrumental (gold) and model (green) data show agreement in terms of both the mean local PSD (dashed) and the PSD of the regional means surface temperature (solid) as the CO\textsubscript{2}-congruent variability dominates. After CO\textsubscript{2}-detrending, the instrumental (orange) and model data (cyan) still agree for the mean local PSD, dominated by local variability, but not for the PSD of the regional mean surface temperature over mainly land  regions. For the latter case, the agreement is better with the internal variability of the models (gray). This highlights differences in spatial structure, with ESDOF increasing with timescale in the absence of external forcing, especially over land.}
    \label{Fig:6Regions}
\end{figure}

\begin{figure}
    \centering
    \includegraphics[width=0.75\textwidth]{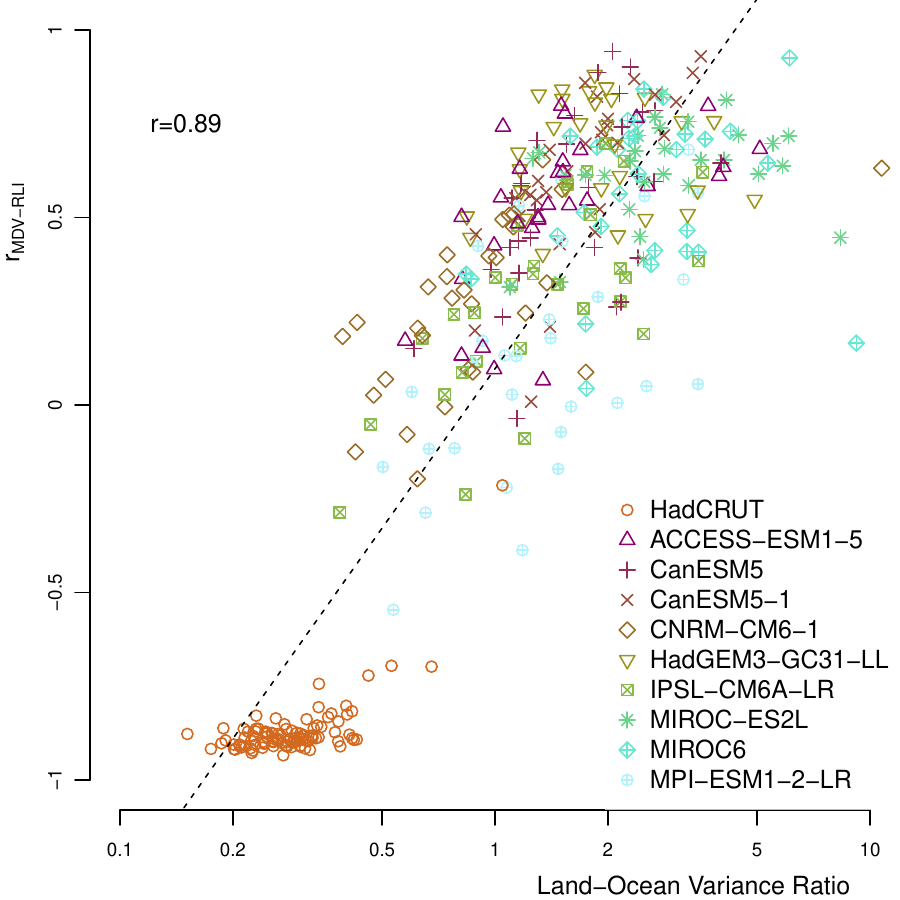}
        \caption{The relationship between $r_{{\sigma^2}_{MD},RLI}$ (the correlation between the multidecadal variance of the CO\textsubscript{2}-detrended fields and RLI) and the multidecadal land-ocean variance ratio between Central Asia and the North Atlantic. The strong correlation between these two metrics (top left) indicates that  $r_{{\sigma^2}_{MD},RLI}$ effectively captures the land-ocean contrast in multidecadal variability, suggesting that it is a robust indicator for assessing regional differences in climate variability between continental and oceanic regions.}
    \label{Supp_Fig_LO_r}
\end{figure}

\begin{figure}
    \centering
    \includegraphics[width=\textwidth]{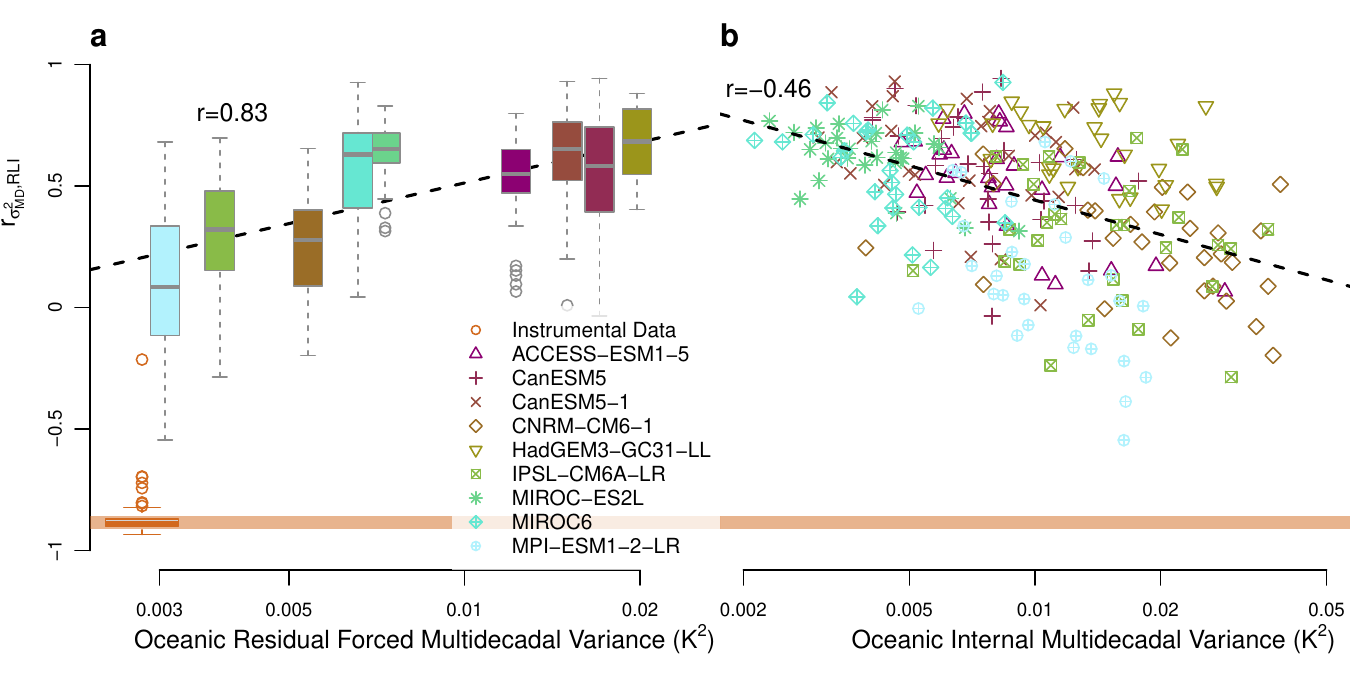}
        \caption{Relationship between the land-ocean contrast (quantified by $r_{{\sigma^2}_{MD},RLI}$ derived from the CO\textsubscript{2}-detrended fields) and oceanic multidecadal (30-80 year) variance (average of the North Pacific and North Atlantic). \textbf{a}, The correlation $r_{{\sigma^2}_{MD},RLI}$ for each model run is plotted against the oceanic residual forced multidecadal variance, calculated from the CO\textsubscript{2}-detrended single-model ensemble mean. Boxplots represent the spread of  $r_{{\sigma^2}_{MD},RLI}$ estimates for individual model realizations with different  internal variability. The regression line (dashed black) and related correlation value are indicated. \textbf{b}, Same as \textbf{a}, but as a function of the amplitude of the oceanic multidecadal variance calculated from the internal variability of each model realisation after subtracting their single-model ensemble mean. }
    \label{Supp_Fig_4}
\end{figure}


\end{document}